\documentclass[pra,aps,twocolumn,superscriptaddress,a4paper,sort&compress,balancelastpage]{revtex4-1}

\usepackage{amsmath,amssymb,calrsfs}
\usepackage{graphicx}
\usepackage[colorlinks=true,citecolor=blue,urlcolor=blue,linkcolor = blue]{hyperref}  
\usepackage{dcolumn}
\usepackage{bm}
\usepackage{verbatim}
\usepackage{units}
\usepackage{float}
\usepackage{upgreek}

\usepackage[T1]{fontenc}
\usepackage[utf8]{inputenc}

\usepackage[dvipsnames]{xcolor}

\usepackage[normalem]{ulem}

\newcommand{\exc}{\mathrm{exc}}
\newcommand{\TOF}{\mathrm{ToF}}

\newcommand{\TITLE}{Universal dynamics of a turbulent superfluid Bose gas}
\begin{document}

\title{\TITLE}
\author{A.D. Garc\'{i}a-Orozco}
	\affiliation{Instituto de F\'isica de S\~ao Carlos, Universidade de S\~ao Paulo, CP 369, 13560-970 S\~ao Carlos, Brazil.}

\author{L. Madeira}
	\email[Corresponding author: ]{madeira@ifsc.usp.br}
	\affiliation{Instituto de F\'isica de S\~ao Carlos, Universidade de S\~ao Paulo, CP 369, 13560-970 S\~ao Carlos, Brazil.}
	
\author{M.A. Moreno-Armijos}
	\affiliation{Instituto de F\'isica de S\~ao Carlos, Universidade de S\~ao Paulo, CP 369, 13560-970 S\~ao Carlos, Brazil.}

\author{A.R. Fritsch}
	\affiliation{Joint Quantum Institute, National Institute of Standards and Technology, and University of Maryland, Gaithersburg, 20899 Maryland, USA}

\author{P.E.S. Tavares}
	\affiliation{Departamento de F\'isica, Universidade Federal de Minas Gerais, CP 702, 31270-901 Belo Horizonte, Brazil}

\author{P.C.M. Castilho}
	\affiliation{Instituto de F\'isica de S\~ao Carlos, Universidade de S\~ao Paulo, CP 369, 13560-970 S\~ao Carlos, Brazil.}

\author{A. Cidrim}
	\affiliation{Departamento de F\'isica, Universidade Federal de S\~ao Carlos, 13565-905 S\~ao Carlos, Brazil.}
	
\author{G. Roati}
	\affiliation{Istituto Nazionale di Ottica del Consiglio Nazionale delle Ricerche (INO-CNR), 50019 Sesto Fiorentino, Italy.}
	\affiliation{European Laboratory for Non-Linear Spectroscopy (LENS), 50019 Sesto Fiorentino, Italy.}
	
\author{V.S. Bagnato}
	\affiliation{Instituto de F\'isica de S\~ao Carlos, Universidade de S\~ao Paulo, CP 369, 13560-970 S\~ao Carlos, Brazil.}
	\affiliation{Hagler Fellow, Hagler Institute for Advanced Study, Texas A\&M University, College Station, Texas 77843, USA}
	
\date{\today}

\begin{abstract}
We study the emergence of universal scaling in the time-evolving momentum distribution of a harmonically trapped three-dimensional Bose-Einstein condensate, parametrically driven to a turbulent state.
We demonstrate that the out-of-equilibrium dynamics post excitation is described by a single function due to nearby non-thermal fixed points. The observed behavior connects the dynamics of a quantum turbulent state to several far-from-equilibrium phenomena. We present a controllable protocol to explore universality in such systems, obtaining the associated scaling exponents. Our experimental results thus offer a promising route to investigate the complex dynamics of the quantum turbulent regime under a novel perspective.
\end{abstract}

\pacs{}
\keywords{}

\maketitle

\section{Introduction}

Understanding how closed many-body quantum systems relax and thermalize when initially prepared far from equilibrium is one of the fundamental questions in modern physics. The topic is relevant to many areas of research, from cosmology \cite{Kofman1994} to high-energy physics \cite{Berges2008}. However, despite intensive studies, many questions are still open. Experiments based on ultracold trapped atoms allow for a precise control and direct observation of their quantum dynamics, accelerating the progress in this direction \cite{Nicklas2015,Prufer2018,Eigen2018,Erne2018,Glidden2021,Galka2022,Madeira2020d,Madeira2022}. At the same time, this experimental approach is boosted by theoretical models that provide a framework to describe this complex phenomenology \cite{Scheppach2010,Karl2013,Berges2015}. In particular, it has been recently proposed that out-of-equilibrium quantum systems can be categorized into classes with universal dynamical behavior, in analogy to universality arising from thermal-fixed points in theories of phase transition \cite{Nowak2013,Schmidt2012}. In this dynamical counterpart, however, universality emerges due to the presence of so-called non-thermal fixed points (NTFPs) -- metastable states of the perturbed quantum many-body system. At the vicinity of these points, far-from-equilibrium systems show no traces of their initial conditions and have their dynamical evolution characterized by only a few parameters \cite{Schmidt2012}. These ideas have successfully described many different out-of-equilibrium phenomena in a generalized manner~\cite{Nowak2012,Nowak2014, Orioli2015,Schmied2018,Chantesana2019,Berges2008,Scheppach2010}.

A paradigmatic example of far-from-equilibrium dynamics for which such universal description is predicted to hold is the quantum turbulent regime in quantum fluids~\cite{Scheppach2010}. Quantum turbulence arises when many quantum vortices tangle with one another \cite{Henn2009, Thompson2013, Yukalov2015,Tsatsos2016, Madeira2020, Madeira2020b} and also when non-linear density waves combine randomly \cite{Nazarenko2011}. A distinctive hallmark of the turbulent regime is the emergence of an energy cascade that corresponds to an atomic momentum distribution described by a power-law over a certain range of wave numbers \cite{Thompson2013,Navon2016}. This cascade mechanism is related to a non-dissipative, self-similar energy transfer between length scales and it reflects the nonlinear dynamics of a turbulent regime \cite{Thompson2013,Tsatsos2016,Tsubota2017}.
There are intrinsic difficulties to identify and characterize quantum turbulence in trapped Bose-Einstein condensates (BECs) based on the power-law behavior alone \cite{Tsatsos2016,Madeira2020}, hence alternatives have been proposed, such as particle and energy fluxes \cite{Baggaley2014,Navon2019,Orozco2020,Marino2020} and entropy related quantities \cite{Madeira2020c}.

In this work, we report the observation of universal dynamics of a far-from-equilibrium three-dimensional (3D), harmonically trapped $\mathrm{^{87}Rb}$ BEC, which is driven to reach a turbulent regime \cite{Henn2009,Thompson2013}. By performing a scaling analysis of the time-evolving momentum distribution $n(k,t)$, we identify a self-similar and universal behavior. This can be characterized by a single universal function, with time and space rescaled by characteristic exponents, $\alpha$ and $\beta$.
We also verify that three different excitation amplitudes lead to the same scaling, suggesting a universal behavior even for distinct initial conditions. In this context, the exponents we extract imply a direct particle cascade, not yet reported in other systems. Our observations in such a distinct scenario corroborate the generality of universal dynamics near NTFPs.

This work is structured as follows. We provide a brief description of the experimental procedure in Sec.~\ref{sec:experimental}. In Sec.~\ref{sec:outofeq}, we present the out-of-equilibrium momentum distributions we obtain. These are analyzed under the concept of far-from-equilibrium states close to NTFPs in Sec.~\ref{sec:universal}. In Sec.~\ref{sec:aniso} we comment on the impact of anisotropy in our findings. Section~\ref{sec:projected} connects the results we obtained using a two-dimensional projection of the atomic cloud to the reconstructed three-dimensional system. Finally, we present our conclusions in Sec.~\ref{sec:conclusion}.

\section{Experimental setup}
\label{sec:experimental}

Our experiment begins with $\mathrm{^{87}Rb}$ BECs having $N = \unit{3.5(3) \times 10^{5}}$ atoms in the $\vert f=2,m_{f} = 2 \rangle$ internal state in a Quadrupole-Ioffe configuration (QUIC) magnetic trap characterized by the frequencies $\omega_{r}/ 2\pi = \unit[130.7(8)]{Hz}$ and $\omega_{x} / 2\pi = \unit[21.8(2)]{Hz}$. The initial equilibrium BEC has a condensate fraction of $70(5)\%$, chemical potential $\mu_0/k_{B} = \unit[124(5)]{nK}$ and healing length $\xi_0 = \unit[0.15(2)]{\mu{m}}$.

\begin{figure}[!htb]
	\centering
	\includegraphics[width=0.9\linewidth]{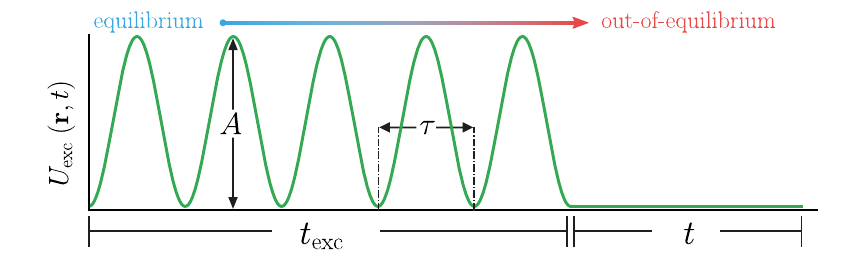}
	\caption{A controllable time-varying magnetic potential $U_\exc$ with amplitude $A$ drives the BEC to an out-of-equilibrium state. The driving potential acts during an excitation time $t_{\exc}$, corresponding to 5 driving periods $\tau$, and then it is turned off. The system evolves in the trap for a holding time $t$, during which there is no external energy input.}
	\label{fig:exc}
\end{figure}

To drive the BEC out of equilibrium we superimpose to the QUIC potential a controllable time-varying magnetic field gradient that creates a potential $U_{\exc}(\mathbf{r},t)= A\big[1-\cos{(\Omega t)}\big]x'/\ell_x$,
where $\ell_x=42\mu$m is a parameter related to the potential acting on the BEC cloud along the coordinate $x$ of the trap, as depicted in Fig.~\ref{fig:exc}.
This is produced by an additional pair of coils in an anti-Helmholtz configuration~\cite{Henn2009} rotated by a small angle ($\approx 5^\circ$) with respect to the principal axis of the QUIC trap. The prime in the $x'$ coordinate indicates that it is to be calculated in the rotated frame.
The application of $U_{\exc}$ corresponds to an effective 3D rotation and distortion of the original trap shape~\cite{Henn2009Vortices}. We verified that our parametric drive couples the dipole mode to shape excitations, such as quadrupolar and scissor modes~\cite{Madison2001}.

In the experiment, we apply a continuous oscillating drive with frequency $\Omega/2\pi = \unit[132.8]{Hz}$, during a time $t_{\exc} = 5\tau$, where $\tau = 2\pi/\Omega$, and the amplitude $A$ is tuned, ranging from from $0$ to $2.8\,\mu_0$. Using this protocol, we thus drive a BEC in initial thermal equilibrium to a far-from-equilibrium state. After the excitation is turned off, the gas is then let to evolve in the trap for variable holding times $t$, where the universal scaling dynamics occurs.


To probe the state of the gas after a time $t$, we turn off the trap potential and take absorption images following the ballistic expansion of the cloud after a time of flight (ToF) of $t_{\TOF}=\unit[30]{ms}$.
The atoms are detected using standard absorption imaging, which gives access to the density distribution in ToF, $n(r)$. During the expansion the momentum of the particles is approximately conserved, thus the density distribution after expansion converges to the in situ momentum distribution of the cloud. Therefore, $n(k)$ is obtained from $n(r)$ by defining $k \equiv mr/(\hbar t_{\TOF})$.

The ToF technique requires a kinetic-energy-dominated state to accurately provide the in situ momentum distribution of the cloud, such is the case of a turbulent state \cite{Caracanhas2013}. In practical terms, this can be achieved for $t_\TOF$ sufficiently larger than $mR/(\hbar k)$, where $R$ is the in situ cloud size. The validity of this method has been extensively discussed in the literature, and this technique has been used successfully to obtain the momentum distribution of turbulent trapped BECs in previous works \cite{Thompson2013,Navon2016}.

\begin{figure}[!htb]
\begin{center}
    \includegraphics[scale=1]{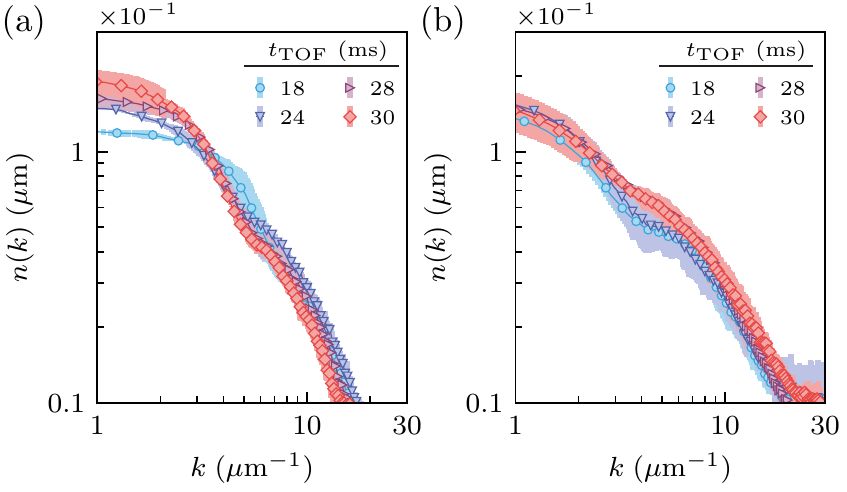}
     \caption{Momentum distributions changing the time of flight ($t_\TOF$) for (a) a quasi-pure BEC and (b) an out-of-equilibrium BEC. The momentum distribution converges rapidly for high-$k$. The $n(k)$ curves overlap for $t_{\TOF}\geqslant \unit[28]{ms}$, which is consistent with $mR/(\hbar k)$. The shaded regions correspond to the experimental uncertainties.}
    \label{fig:nk_TOF}
    \end{center}
\end{figure}

We assess the validity of this method by changing the expansion time and comparing the resulting distributions, as shown in Fig.~\ref{fig:nk_TOF} for (a) a quasi-pure BEC and (b) an out-of-equilibrium BEC. In both panels of Fig.~\ref{fig:nk_TOF}, after $t_{\TOF}=\unit[28]{ms}$ all $n(k)$ curves converge to the same distribution within the experimental uncertainty.

\section{Out-of-equilibrium distributions}
\label{sec:outofeq}

\begin{figure}[!htb]
\begin{center}
\includegraphics[width=0.8\linewidth]{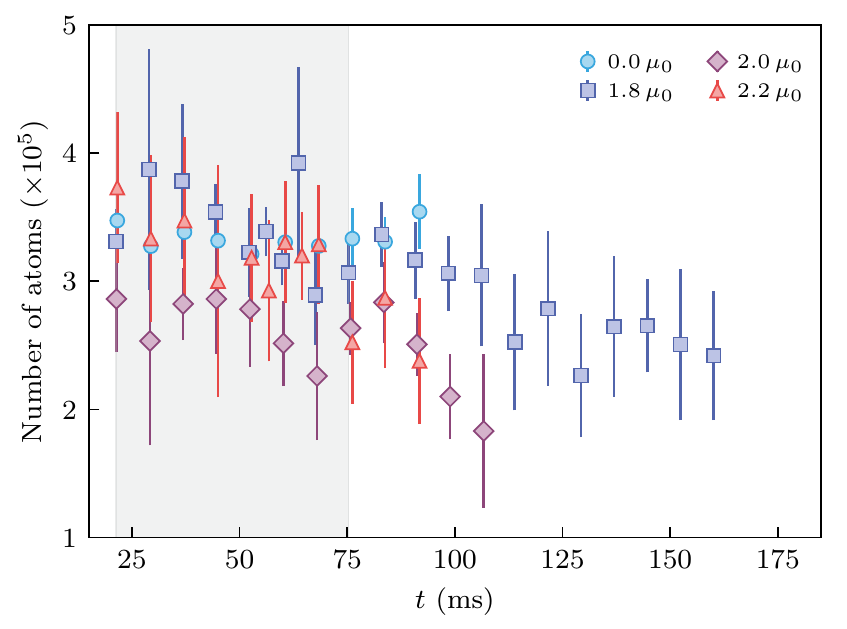}
\caption{
Number of atoms as a function of time for the condensate in equilibrium ($A=0$), and three different excitation amplitudes, $A = 1.8$, 2.0, and 2.2$\mu_0$.
The shaded region corresponds to the time window when the universal scaling is observed.}
\label{fig:number}
\end{center}
\end{figure}

The momentum distribution $\tilde{n}(k, t)$ is obtained from the two-dimensional projection of the cloud. 
Averaging and appropriately transforming position ($x$ and $y$) into momentum ($k_x$ and $k_y$) as in $x =\hbar t_\TOF k_x / m$ (and similarly for $y$), we obtain the projections of the \textit{in-situ} momentum distributions $\tilde{n}(k)$,
$k=(k_x^2+k_y^2)^{1/2}$, for each instant $t$ of the evolution time after the excitation.
The normalized momentum distribution is given by $n(k,t)=\tilde{n}(k, t)/N(t)$, where $N(t)$ is the total number of atoms at the holding time $t$.
Figure~\ref{fig:number} depicts $N(t)$ for an atomic cloud in equilibrium, and also for finite excitation amplitudes.

\begin{figure}[!htb]
	\centering
	\includegraphics[width=\linewidth]{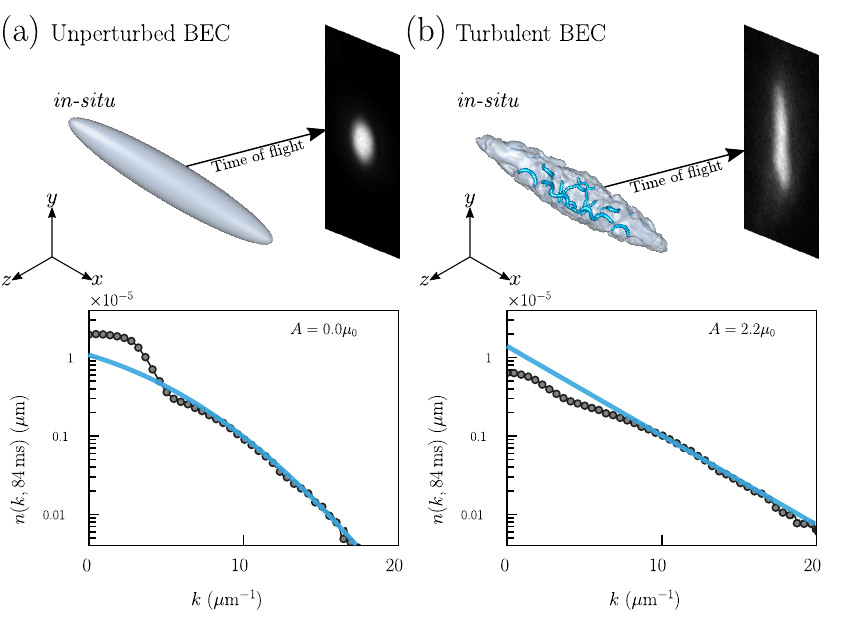}
	\caption{Momentum distributions for $t= \unit[84]{ms}$ integrated along $\hat{z}$. (a) Without excitation ($A=0.0 \mu_0$), the BEC is characterized by a central peak for low-momenta and a Gaussian (blue line) distribution for the thermal component. (b) When the excitation amplitude is large enough, in this case $A=2.2\mu_0$, a far-from-equilibrium turbulent regime is reached, and the distribution for atoms with higher momentum shows the transition to a clear non-Gaussian distribution.}
	\label{fig:profiles}
\end{figure}

In Figs.~\ref{fig:profiles}(a) and (b) we show the integration of in-plane density profiles, resulting in on-axis distributions, obtained from absorption images after $t= \unit[84]{ms}$ for a BEC both in equilibrium and driven by an amplitude $A=2.2\,\mu_0$, respectively. 
In Fig.~\ref{fig:profiles}(a), the usual bimodal $n(k)$ of a BEC in thermal equilibrium is displayed, with a high-density peak region corresponding to the occupation of a condensate fraction (the well-known Thomas-Fermi regime) and the Gaussian distribution (blue line) due to a thermal component. For a BEC driven to a turbulent state, the system exhibits a broadened momentum distribution.
The central peak is shrunk as a consequence of particle removal from lower momenta. At the same time, the tails show a departure from the Gaussian shape to an exponential dependence, as depicted in Fig.~\ref{fig:profiles}(b).
The latter is a signature of deviation from equilibrium caused by the driving potential.

\begin{figure}[!htb]
	\begin{center}
	\includegraphics[width=\columnwidth]{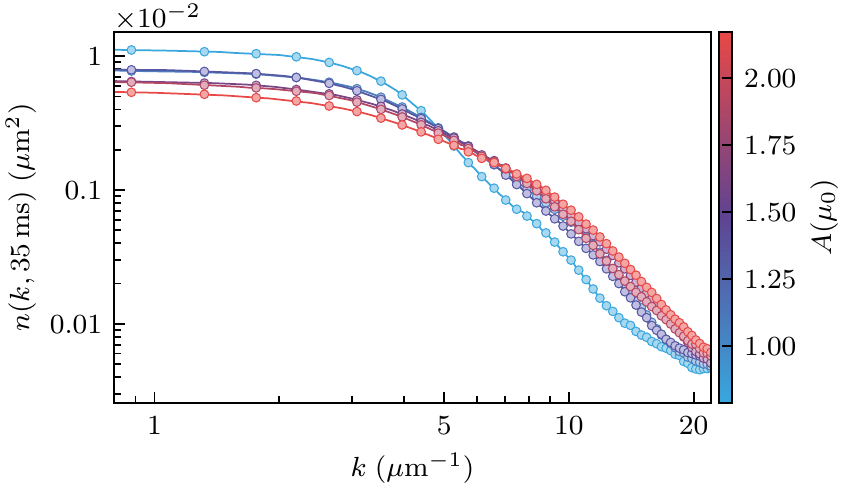}
	\caption{Momentum distributions $n(k,t\text{=35 ms})$ for different excitation amplitudes. The establishment of a turbulent state is supported by the power-law behavior in the momentum range $\unit[10]{\mu m^{-1}} \leqslant k \leqslant \unit[17]{\mu m^{-1}}$.}
	\label{fig:thold30}
	\end{center}
\end{figure}

In Fig.~\ref{fig:thold30}, we show typical momentum distributions for a fixed time $t=35$ ms and different excitation amplitudes $A$ after performing an angular average over the in-plane momentum shells of radius $k$.
Hereafter we chose not to include error bars in the figures concerning the momentum distributions for clarity. Typically they are of the order of $10\%$ and do not exceed $15\%$.
For fixed excitation times, as $A$ increases, population at higher momenta grows. A power-law behavior appears in the momentum distribution, $n(k) \propto k^{-3.1(1)}$, over a $k$-range of $\unit[10]{\mu m^{-1}} \leqslant k \leqslant \unit[17]{\mu m^{-1}}$ and within the time window $\unit[20]{ms}\lesssim t \lesssim \unit[70]{ms}$ for excitation amplitudes of $1.8\mu_0 \leqslant A\leqslant 2.2\mu_0$. This signals an energy cascade and thus the emergence of a turbulent state in the sample under consideration, as observed in other previous experiments~\cite{Thompson2013}. After $t \approx \unit[100]{ms}$, the transient turbulent state decays, thus relaxing toward thermalization. For $A\gtrsim 2.4\,\mu_0$, we verified that the final state is a thermal gas, indicating that the drive has injected enough energy to fully deplete the condensate.

\section{Universal scaling}
\label{sec:universal}

\begin{figure*}[!htb]
    \begin{center}
    \includegraphics[width=0.7\linewidth]{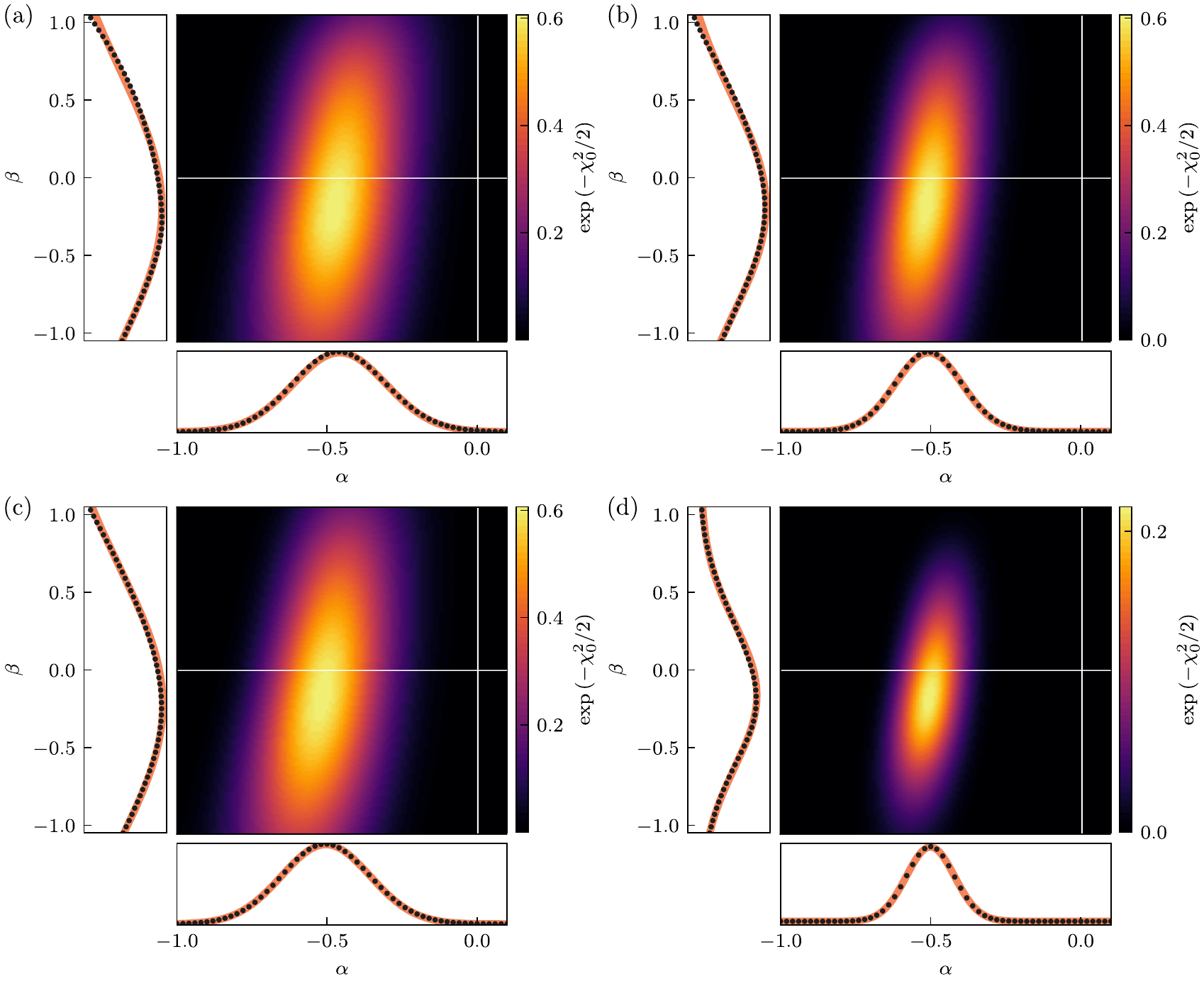}
    \caption{Likelihood functions $L_A(\alpha,\beta)$ for three different excitation amplitudes, $1.8\mu_0$ (a), $2.0\mu_0$ (b), and $2.2\mu_0$ (c), and the product of all likelihood functions (d). For panels (a)-(c) we divide the function $\chi^2(\alpha,\beta)$ by its minimum value, which we denote by $\chi_0^2(\alpha,\beta)$, so that the maximum value of $L_A(\alpha,\beta)$ is the same for all excitation amplitudes. This choice does not affect the procedure to obtain the exponents, since the Gaussian fits to the marginal likelihood functions, shown in the left and bottom panels of each figure, are just multiplied by a constant. Note that the combined likelihood function (d) is much more localized than the individual excitation amplitudes, as we would have expected.
    }
    \label{fig:likelihood}
    \end{center}
\end{figure*}

Once the system is driven out of equilibrium, we cease the parametric excitation and let the turbulent state enter a relaxation and thermalization dynamics, and the temporal evolution of the system is recorded.
We now analyze these distributions under the concept of universal dynamics exhibited by far-from-equilibrium quantum systems close to NTFPs~\cite{Erne2018,Prufer2018,Chantesana2019,Berges2015,Schmidt2012,Madeira2022}. It has been proposed \cite{Orioli2015} that far-from-equilibrium closed systems that belong to a certain universality class should exhibit their universal character through the distribution $n (k, t)$, which scales in time and momentum following the form
\begin{equation}
n \left( k,t \right) = \left( \frac{t}{t_0} \right)^{\alpha} F \left[ \left( \frac{t}{t_0} \right)^\beta k \right],
\label{eq:universal}
\end{equation}
\noindent with $t_0$ being an arbitrary reference time within the period in which $n \left(k, t\right)$ shows scaling properties. The $\alpha$ and $\beta$ exponents must be universal and independent of the initial conditions of the system. This being true, it shows that, over a certain momentum range, the distribution $n \left(k, t\right)$ of the decaying turbulent system depends on space and time only through a single universal function $F\left(k \right)$. 

\begin{figure*}[!htb]
	\begin{center}
	\includegraphics[width=0.8\linewidth]{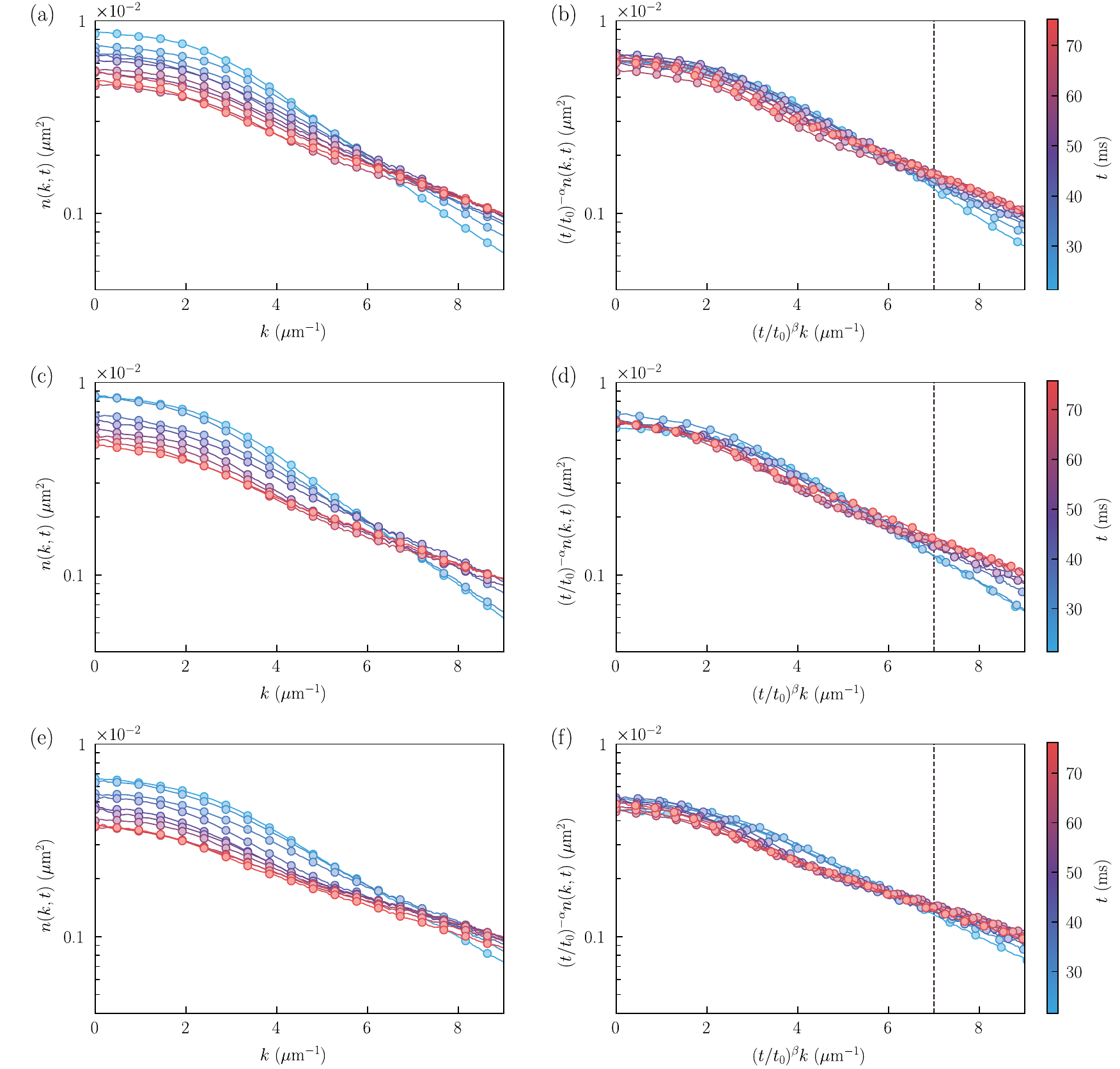}
    \caption{(a,c,e) Momentum distributions of turbulent states for excitation amplitudes of $A = 1.8$, 2.0, and 2.2$\mu_0$ and different holding times $t$. (b,d,f) After the rescaling, we see that all distributions collapse into a single curve, signaling universal dynamical behavior in the turbulent regime. The exponents characterizing this universality class are $\alpha = -0.50(8)$ and $\beta = -0.2(4)$. The vertical dashed line indicates the cutoff $k_s$ of the infrared region, where the universal scaling following Eq.~(\ref{eq:universal}) is observed.}
	\label{fig:universality}
	\end{center}
\end{figure*}

We extracted the universal scaling exponents in Eq.~(\ref{eq:universal}) following closely the procedure adopted in Ref.~\cite{Erne2018}, which was done in the context of a far-from-equilibrium one-dimensional Bose gas emerging from a strong cooling quench of a three-dimensional system. We minimized the function
\begin{equation}
\chi^2(\alpha,\beta)=\frac{1}{N_t^2}\sum_{t=t_1}^{t_{N_t}}\sum_{t_0=t_1}^{t_{N_t}} \chi^2_{\alpha,\beta}(t,t_0),
\end{equation}
where we average both the times $t$ and reference times $t_0$ over all the $N_t$ holding times $\{t_1,\cdots,t_{N_t}\}$. The function $\chi^2_{\alpha,\beta}(t,t_0)$ is given by
\begin{equation}
\label{eq:local}
\chi^2_{\alpha,\beta}(t,t_0)=\int_{k_i}^{k_f} dk \frac{[(t/t_0)^\alpha n((t/t_0)^\beta k,t_0)-n(k,t)]^2}{\sigma((t/t_0)^\beta k,t_0)^2+\sigma(k,t)^2},
\end{equation}
where $\sigma(k,t)$ corresponds to the standard deviation of the mean. The quantities
$n(k,t)$ and $\sigma(k,t)$ are normalized by the total number
of atoms, $n(k,t)=\tilde{n}(k,t)/N(t)$ and $\sigma(k,t)=\tilde{\sigma}(k,t)/N(t)$. The integration in Eq.~(\ref{eq:local}) is done over a $k$-range $[k_i,k_f]$. We chose $k_i$ to be the lowest value available from the experimental data, and $k_f$ was varied to guarantee that the results are independent of our choice.
We found that the exponents are insensitive to variations of $\delta k\approx\unit[0.5]{\mu{m}^{-1}}$ around $k_f=\unit[10]{\mu{m}^{-1}}$ and $\delta t\approx\unit[5]{ms}$ (i.e., increasing or decreasing the initial and final times of the scaling window), within reasonable limits.

The values of $\alpha$ and $\beta$ are estimated through a likelihood function,
\begin{equation}
L_A(\alpha,\beta)=\exp\left[-\frac{1}{2}\chi^2(\alpha,\beta)\right],
\end{equation}
where the subscript $A$ is a label for the different excitation amplitudes. In Fig.~\ref{fig:likelihood}(a)-(c) depict these functions for $A=1.8$, $2.0$, and $2.2\mu_0$, respectively. The values of the exponents and their uncertainties are determined from a Gaussian fit of the the marginal-likelihood functions,
\begin{eqnarray}
L_{\alpha,A}(\alpha)=\int d\beta \ L_A(\alpha,\beta),\nonumber\\
L_{\beta,A}(\beta)=\int d\alpha \ L_A(\alpha,\beta),
\end{eqnarray}
which are shown in the bottom and left panels of Figs.~\ref{fig:likelihood}(a)-(c).
Since these amplitudes produced essentially the same exponents, the values of $\alpha$ and $\beta$ and their uncertainties are estimated from the combined likelihood function,
\begin{equation}
L(\alpha,\beta)= \prod_A L_{A}(\alpha,\beta),
\end{equation}
shown in Fig.~\ref{fig:likelihood}(d).

\begin{figure*}[!htb]
    \begin{center}
	\includegraphics[width=0.8\linewidth]{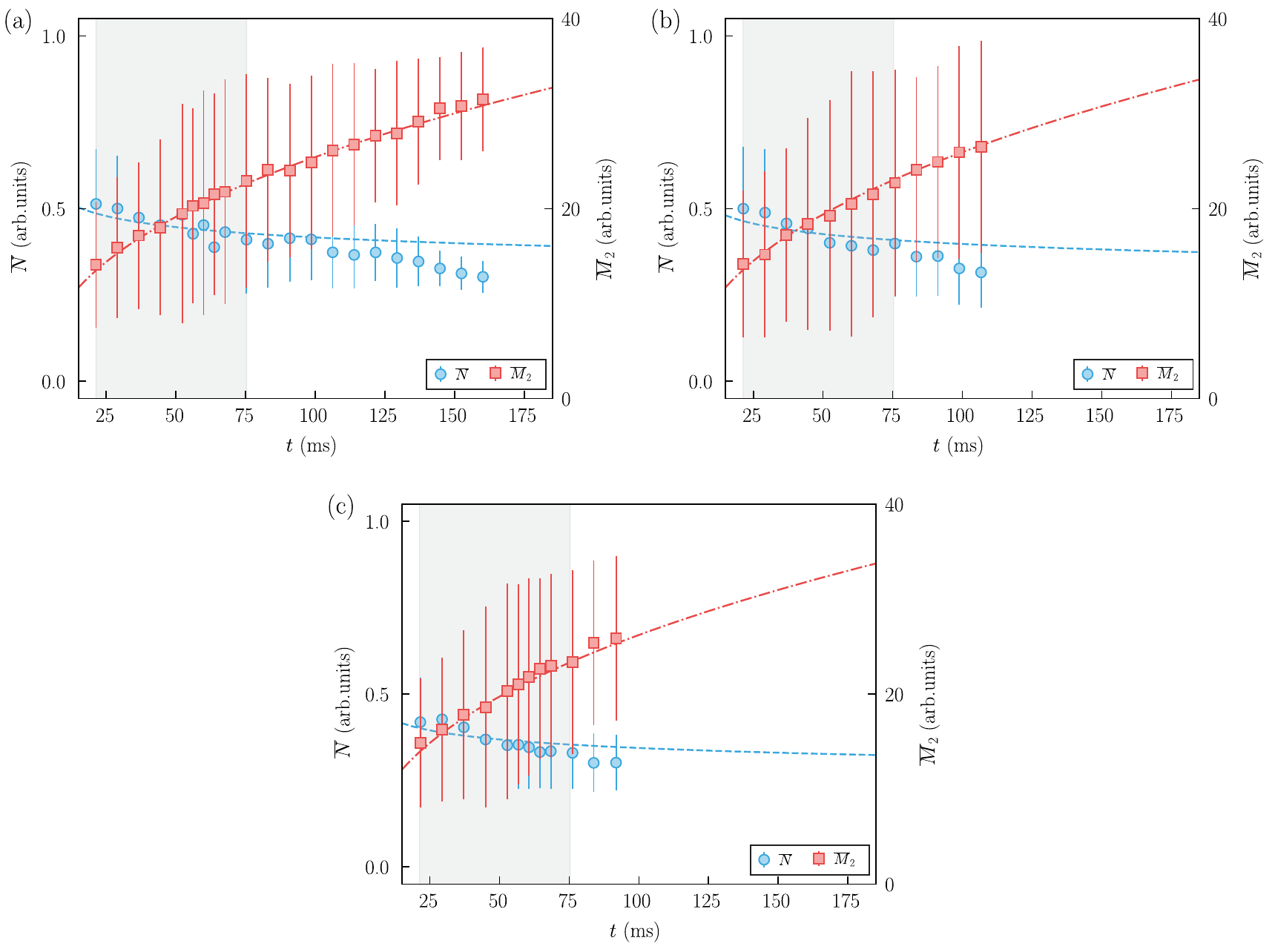}
	\caption{Global quantities, as defined in Eqs.~(\ref{eq:nbar}) and (\ref{eq:m2}), for the excitation amplitudes of $A = 1.8$, 2.0, and 2.2$\mu_0$, panels (a,b,c), respectively.
The shaded region corresponds to the time window in which universal behavior is observed. The left axis is the total number of particles $\bar{N}$ over the universality range ($k_s \leqslant \unit[7]{\mu m^{-1}}$) as a function of $t$.  The mean kinetic energy $\overline{M}_{2}$, right axis, increases with time over the same universal range. The curves follow the theoretical predictions, namely $\bar{N}\propto t^{\alpha-2\beta}$ and $\bar{M}_2\propto t^{-2\beta}$.}
	\label{fig:conservation}
	\end{center}
\end{figure*}

The temporal evolution of the system is recorded for different excitation amplitudes, $A=1.8$, 2.0, and 2.2$\mu_0$, as shown in Figs.~\ref{fig:universality}(a,c,e). As we wait longer, the distributions evolve, promoting more population from low to high momentum values.
The rescaling of the different distributions is provided in Figs.~\ref{fig:universality}(b,d,f) for the infrared region ($k\leqslant k_\text{s}=\unit[7]{\mu m^{-1}}$), showing that all data for different times fall onto a single curve, with scaling exponents $\alpha = -0.50(8)$ and $\beta =-0.2(4)$.

This self-similar evolution for the infrared momentum range is related to the transport of particles in our closed system during the selected time window. Two global quantities can be defined, 
\begin{eqnarray}
\label{eq:nbar}
\overline{N}
&=&\int\limits_{|k|\leqslant\left(\frac{t}{t_0}\right)^{-\beta}k_s} d^dk \ n(k,t)
\propto \left(\frac{t}{t_0}\right)^{\alpha-d\beta},\\
\label{eq:m2}
\overline{M}_2&=&\int\limits_{|k|\leqslant\left(\frac{t}{t_0}\right)^{-\beta}k_s} d^dk \ k^2\frac{n(k,t)}{\overline{N}(t)} \propto \left(\frac{t}{t_0}\right)^{-2\beta},
\end{eqnarray}
where $k_s$ defines the high-momentum cutoff for the scaling region.
We determined this characteristic scale by comparing each of the scaled momentum distributions with their average.
We considered the value $k_s=\unit[7.0]{\mu{m}^{-1}}$, which corresponds to a difference greater than one standard deviation.

Self-similarity in this case requires that the particle number $\bar{N}$ over that range of scaling is conserved.
Figures~\ref{fig:conservation}(a-c) show that there is a slight decrease when the average number of particles is evaluated in the scaling region $k\leqslant k_\text{s}$ ($\bar{N}\propto t^{-0.1}$ for $A = 1.8$ $\mu_0$, for example).
Since its time dependence does not produce abrupt changes, we can consider it approximately constant in the scaling region.
As a consequence of the particle number being approximately conserved in the dynamics, the average kinetic energy should increase over this same range, following $\bar{M}_{2}\propto t^{-2\beta}$~\cite{Erne2018}. We indeed observe this buildup of the energy in the scaling region, as can be verified in Figs.~\ref{fig:conservation}(a-c).

\section{Anisotropy of the atomic cloud}
\label{sec:aniso}

\begin{figure}[!htb]
	\begin{center}
	\includegraphics[width=0.9\linewidth]{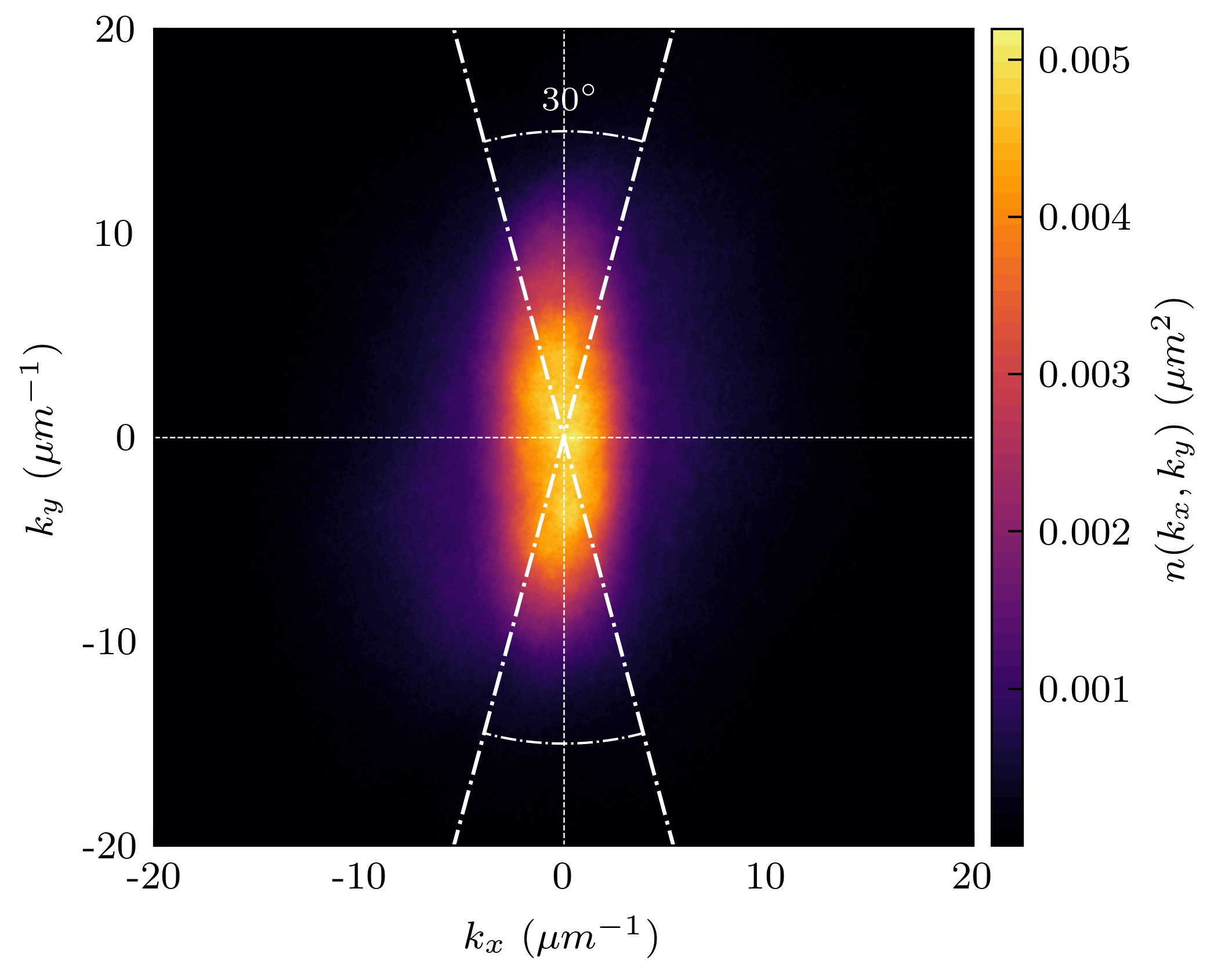}
	\caption{
Momentum distribution for an excitation amplitude of $A=1.8\mu_0$ and a holding time of $t=\unit[60]{ms}$, which is clearly anisotropic. Momentum distributions for other values of the excitation amplitude and holding times are qualitatively similar. The lines, given by $k_y/k_x=\tan (90^{\circ}\pm 15^{\circ})$, denote two 30$^{\circ}$ angles centered around the major axis of the expanded cloud.
}
	\label{fig:cloud_major}
	\end{center}
\end{figure}

The momentum distributions reported in this work are obtained from the expansion of cigar-shaped BECs, which are anisotropic. In Fig.~\ref{fig:cloud_major}, we show a typical momentum distribution as a function both $k_x$ and $k_y$.
The universal scaling of Eq.~(\ref{eq:universal}) assumes isotropy of the momentum distribution, i.e., it depends only on $k=(k_x^2+k_y^2)^{1/2}$. Thus, the angular averaging procedure reported in this work has to be justified.

In this section, we examined the momentum distributions computed with an angular average only in the regions close to the major axis of the expanded cloud (which corresponds to the minor axis of the \textit{in-situ} cloud). We chose an angular aperture of 30$^\circ$ around the major axis, which is denoted by the $k_y/k_x=\tan (90^{\circ}\pm 15^{\circ})$ lines in Fig.~\ref{fig:cloud_major}. In this region, the distribution is approximately isotropic. The comparison between the results obtained using only this portion and the entire momentum distribution can shed light on the impact of anisotropy in our findings.

\begin{figure*}[!htb]
	\begin{center}
	\includegraphics[width=0.8\linewidth]{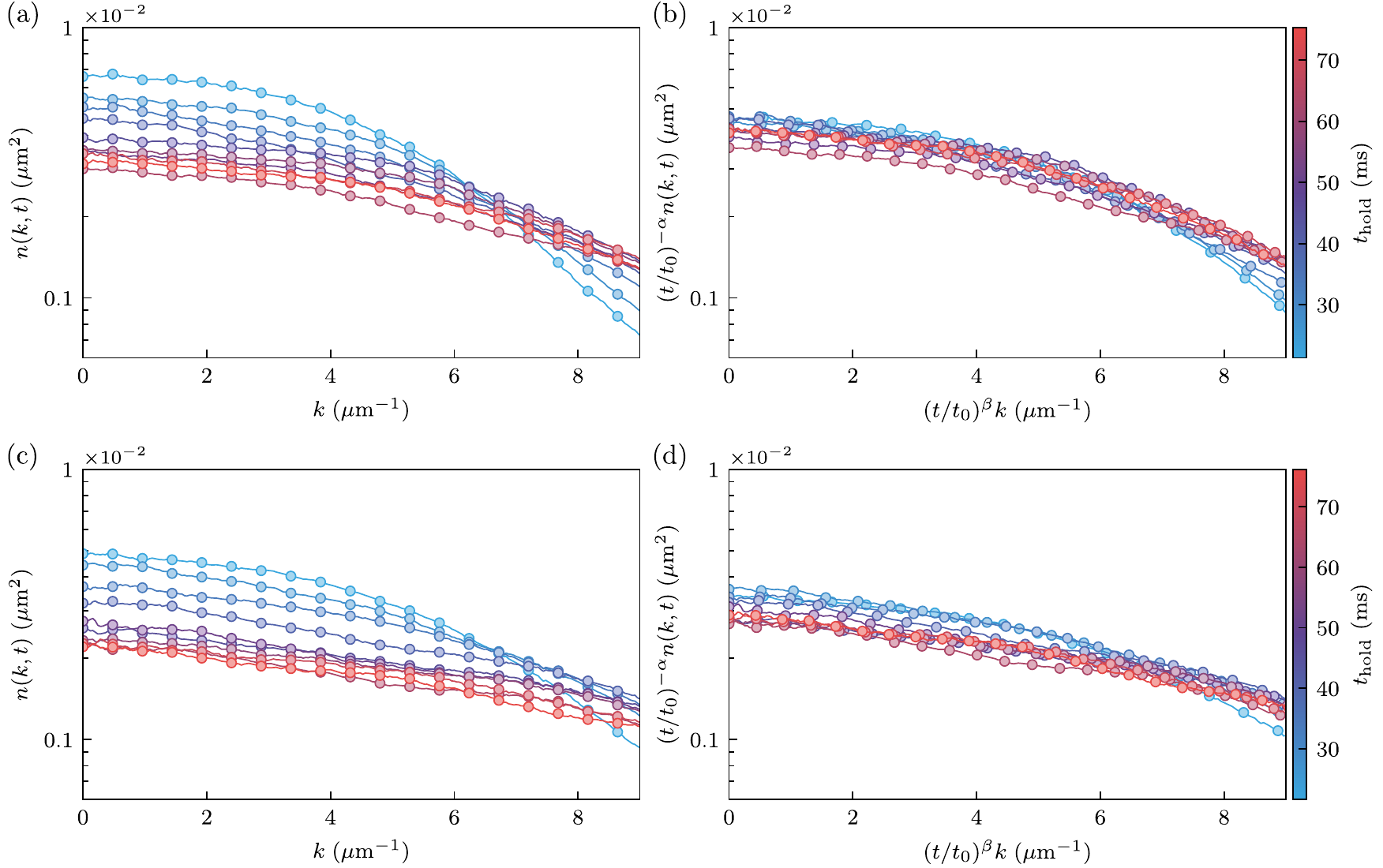}
	\caption{
(a,c) Momentum distributions for the lowest and highest excitation amplitudes considered, $A=1.8$ and $2.2\mu_0$, respectively, obtained with the angular average over only the regions close to the major axis of the expanded cloud. (b,d) Corresponding scaled momentum distributions with the exponents $\alpha=-0.50$ and $\beta=-0.2$.
}
	\label{fig:major}
	\end{center}
\end{figure*}

The momentum distributions for different excitation amplitudes obtained with this restricted angular average are plotted in Figs.~\ref{fig:major}(a,c). Their normalization is chosen such that $\int dk \ k \ n(k)=1$ to make comparisons easier with the momentum distributions reported in the other sections. In Figs.~\ref{fig:major}(b,d) we employ the same exponents reported in the main text, $\alpha=-0.50$ and $\beta=-0.2$, to scale the $n(k)$ according to Eq.~(\ref{eq:universal}).

It is possible to see that, for all cases, the spread of the curves is reduced after the scaling has been applied. Hence anisotropy cannot play a major role in the results reported in this work.
However, it should lead to some corrections since the collapse of all curves into a single universal function in Figs.~\ref{fig:major}(b,d) is not as good as the one for momentum distributions obtained with the angular average over the whole cloud.
The effect should be magnified as we increase the excitation amplitude because the momentum distributions become more elongated, which is confirmed by comparing Figs.~\ref{fig:major}(b) and (d).

\section{Scaling of a projected momentum distribution}
\label{sec:projected}

The procedure described so far involves the $n(k,t)$ obtained from absorption images of the cloud, which correspond to two-dimensional projections of the three-dimensional system. A question that arises is if the exponents obtained through the projections can be related to the ones of the three-dimensional system.

In this section, we employed the subscripts 3D and 2D to explicitly differentiate the three-dimensional momentum distribution and its two-dimensional projection measured in our experiments, respectively.
To avoid a cumbersome notation in all other sections, the quantities correspond to the two-dimensional case, unless stated otherwise.

The Abel transform~\cite{Bracewell1986} of a spherically symmetric function $n_{\rm 3D}(k)$ yields its projection on a plane, $n_{\rm 2D}(k)$. It is an integral transformation which can be written as
\begin{equation}
\label{eq:abel}
n_{\rm 2D}(k) = \int_k^\infty dk' \frac{n_{\rm 3D}(k')k'}{\sqrt{k^{'2}-k^2}}.
\end{equation}
The inverse Abel transform has been successfully used in previous works ~\cite{Thompson2013,Navon2016} to reconstruct the momentum distribution of a three-dimensional cloud from the two-dimensional absorption images.

Let us start with an isotropic (for simplicity) three-dimensional momentum distribution that obeys the universal scaling near a NTFP,
\begin{equation}
n_{\rm 3D} \left( k,t \right) = \left( \frac{t}{t_0} \right)^{\alpha_{\rm 3D}} F_{\rm 3D} \left[ \left( \frac{t}{t_0} \right)^{\beta_{\rm 3D}} k \right].
\end{equation}
Its projection after a time $t$ is given by Eq.~(\ref{eq:abel}),
\begin{equation}
n_{\rm 2D}(k,t) = \int_k^\infty dk' \left(\frac{t}{t_0}\right)^{\alpha_{\rm 3D}} \frac{F_{\rm 3D}\left[\left(\frac{t}{t_0}\right)^{\beta_{\rm 3D}} k'\right] \ k'}{\sqrt{k^{'2}-k^2}}.
\end{equation}
Changing variables and computing the projection at $(t/t_0)^{-\beta_{\rm 3D}}k$,
\begin{eqnarray}
n_{\rm 2D}\left[\left(\frac{t}{t_0}\right)^{-\beta_{\rm 3D}} k,t\right]
&=&
\left(\frac{t}{t_0}\right)^{\alpha_{\rm 3D}-\beta_{\rm 3D}}
\int\limits_{k}^\infty
d\tilde{k}  \frac{F_{\rm 3D}(\tilde{k}) \ \tilde{k}}{\sqrt{\tilde{k}^2-k^2}}
\nonumber\\
&=&\left(\frac{t}{t_0}\right)^{\alpha_{\rm 3D}-\beta_{\rm 3D}}
F_{\rm 2D}(k),
\end{eqnarray}
where the last equality comes from identifying the integral as the Abel transform of the universal function, $F_{\rm 3D}(k)$.
Finally, computing the expression for $n_{\rm 2D}\left(k,t\right)$,
\begin{eqnarray}
n_{\rm 2D}\left(k,t\right)
=\left(\frac{t}{t_0}\right)^{\alpha_{\rm 3D}-\beta_{\rm 3D}}
F_{\rm 2D}\left[\left(\frac{t}{t_0}\right)^{\beta_{\rm 3D}} k\right].
\end{eqnarray}
The expression above is the universal scaling of the projection,
\begin{equation}
n_{\rm 2D} \left( k,t \right) = \left( \frac{t}{t_0} \right)^{\alpha_{\rm 2D}} F_{\rm 2D} \left[ \left( \frac{t}{t_0} \right)^{\beta_{\rm 2D}} k \right],
\end{equation}
provided that we identify
\begin{equation}
\alpha_{\rm 2D}=\alpha_{\rm 3D}-\beta_{\rm 3D} \text{ and } \beta_{\rm 2D}=\beta_{\rm 3D}.
\end{equation}
Furthermore, if we assume $\alpha_{\rm 3D}=3\beta_{\rm 3D}$, which comes from particle conservation in the scaling region, see Eq.~(\ref{eq:nbar}), we obtain particle conservation in the projected system $\alpha_{\rm 2D}=2\beta_{\rm 2D}$, and also the relation between the values of $\alpha$, $\alpha_{\rm 2D}=2\alpha_{\rm 3D}/3$.

These results tell us that the universal scaling of an isotropic three-dimensional momentum distribution survives the projection procedure.
Hence, it is possible to investigate the scaling of a 3D isotropic system by studying only its 2D projection without reconstructing the 3D momentum distribution.

The momentum distributions reported in this work are not isotropic, as discussed in Sec.~\ref{sec:aniso}.
However, we can still investigate if these analytical predictions hold. To this end, we employed the inverse Abel transform~\cite{Bracewell1986} to reconstruct the three-dimensional momentum distributions from their projections,
\begin{equation}
\label{eq:invabel}
n_{\rm 3D}(k,t) = -\frac{1}{\pi}\int_k^\infty dk' \frac{dn_{\rm 2D}(k',t)}{dk'} \frac{1}{\sqrt{k^{'2}-k^2}}.
\end{equation}
Note that there is a derivative of the two-dimensional momentum distribution with respect to the momentum, which has to be taken numerically and introduces noise.

\begin{figure*}[!htb]
	\begin{center}
	\includegraphics[width=\linewidth]{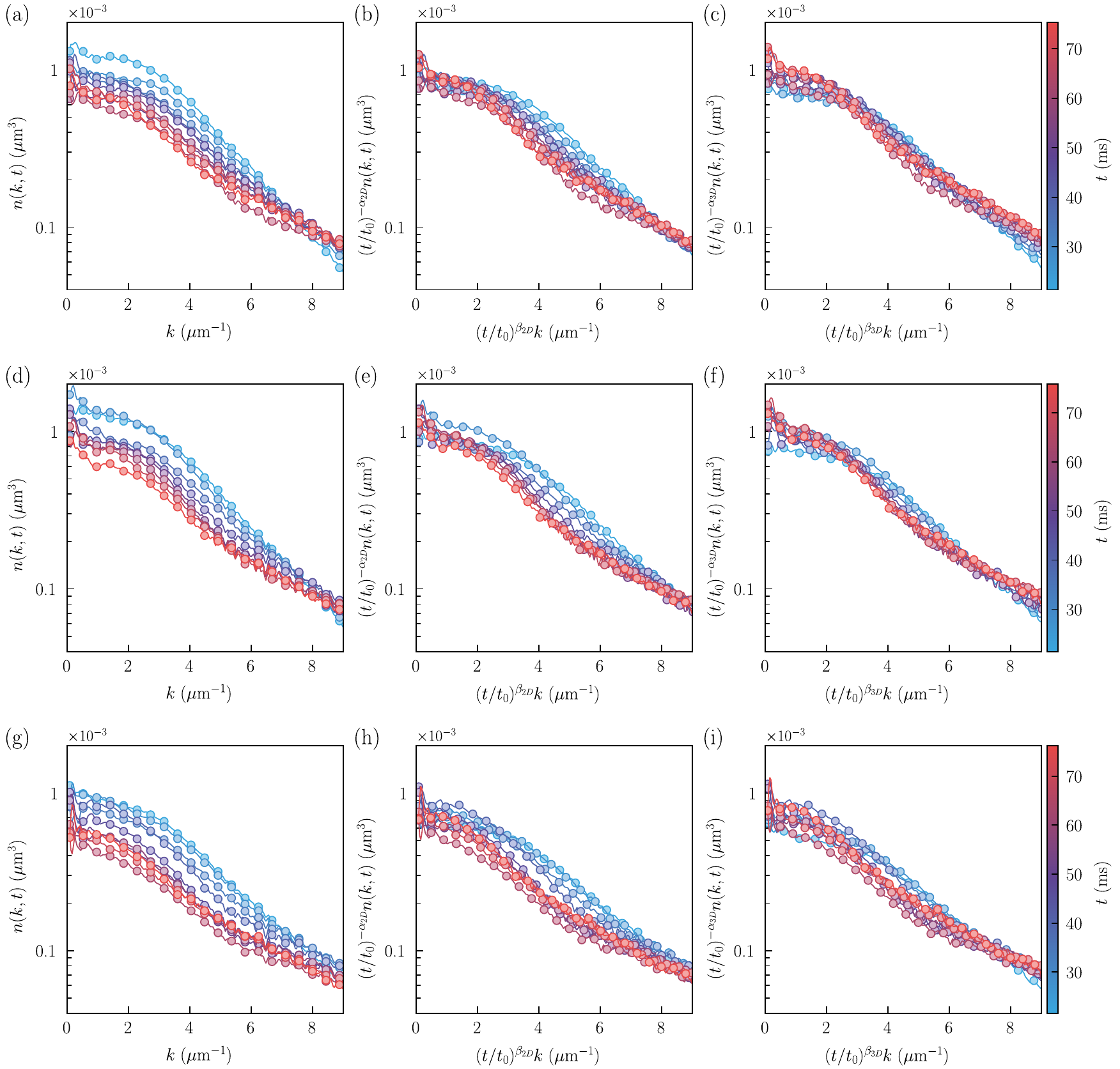}
	\caption{
	(a,d,g) Three-dimensional momentum distributions obtained with the inverse Abel transform, Eq.~(\ref{eq:invabel}), for $A=1.8,$ $2.0$ and $2.2\mu_0$, respectively. (b,e,h) Scaled momentum distributions using the exponents $\alpha_{\rm 2D}=-0.50$ and $\beta_{\rm 2D}=-0.2$. (c,f,i) Scaled momentum distributions using the theoretical prediction for isotropic $n_{\rm 3D}(k)$, $\alpha_{\rm 3D}=3\alpha_{\rm 2D}/2=-0.75$ and $\beta_{\rm 3D}=\beta_{\rm 2D=}-0.2$.
Notice that the collapse into a universal function is much better using the 3D exponents.
}
	\label{fig:abel_sup}
	\end{center}
\end{figure*}

In Figs.~\ref{fig:abel_sup}(a,d,g), we present the reconstruction of the three-dimensional momentum distributions from their projections using the inverse Abel transform.
They are normalized according to $\int dk \ k^2 n_{\rm 3D}(k)=1$ to make comparisons with the other momentum distributions more straightforward.
We attempt the scaling using the exponents calculated with the projections, $\alpha_{\rm 2D}=-0.50$ and $\beta_{\rm 2D}=-0.2$, in Figs.~\ref{fig:abel_sup}(b,e,h), which fails to collapse into a single universal function.
Using instead our prediction for the three-dimensional scaling exponents, $\alpha_{\rm 3D}=3\alpha_{\rm 2D}/2=-0.75$ and $\beta_{\rm 3D}=-0.2$, Figs.~\ref{fig:abel_sup}(c,f,i), the agreement is improved considerably. This is another piece of evidence that the impacts due to the anisotropy of the momentum distribution, although present, do not influence our main results significantly.

\section{Conclusion}
\label{sec:conclusion}

Different excitation amplitudes, $A=1.8\,\mu_0$, $2.0\,\mu_0$, and $2.2\,\mu_0$, and therefore distinct initial states, lead to the same type of distribution with equivalent exponents. They thus belong to the same class of universal dynamics.
For a thermally quenched 3D, isotropic and homogeneous Bose gas, numerical and analytical calculations for the infrared region have predicted the relation $\alpha=\beta d$ (where $d$ is the dimension of the system)~\cite{Orioli2015}. Remarkably, our anisotropic, harmonically trapped system follows the same correspondence for the scaling exponents over a range of amplitudes.

Our exponents, however, do not follow the absolute values predicted in the numerical simulations of \cite{Orioli2015}, namely: $\alpha=1.66(12)$ and $\beta=0.55(3)$. This disagreement is expected for several reasons. First of all, despite being 3D, we have a finite system in an anisotropic trap, imposing inhomogeneity over large spatial scales.
Second, we provide a different route from quenching protocols to reach a far-from-equilibrium state in spinless trapped BECs. Some assumptions in \cite{Orioli2015} such as equally populating momentum states in the initial state preparation certainly does not hold in our case, and may lead the dynamics to different NTFPs. Lastly, we obtain universal scaling exponents with negative signs. This reveals a \textit{direct} particle cascade, which has been observed in perturbative estimates~\cite{Micha2004} of turbulence thermalization. However, this is the opposite of Ref.~\cite{Orioli2015} and recent experiments~\cite{Erne2018,Glidden2021} of after-quench dynamics with ultracold gases.
The condensation process observed in thermally quenched Bose gases, forming a quasi-condensate at intermediate times via particle-conserving transport to lower momenta, is absent here.

Surprisingly, the relation between the scaling exponents $\alpha$ and $\beta$ still retains the information about the system before the projection from the absorption imaging.
Ideally, it would be interesting to implement experimental techniques that allow us to obtain the three-dimensional momentum distribution directly, and not only its in-plane projection as employed in this work. Prospective studies might also focus on connecting the analysis provided here with the concept of the inverse Kibble-Zurek mechanism~\cite{Yukalov2015b}.

The universal scaling of Eq.~(\ref{eq:universal}) relies on two exponents, which we extracted from the data, and a universal function $F(k)$, which we present in Figs.~\ref{fig:universality}, \ref{fig:major}, and \ref{fig:abel_sup}. However, we do not propose a functional form for it, because we would need to know the mechanisms behind the observed turbulence. The question of which type of turbulence is generated by a given excitation is still an open one~\cite{Madeira2020}.
It is our intention to investigate this in detail in future works.

We hope this work serves as motivation to investigate both anisotropy and non-homogeneous densities in far-from-equilibrium systems.
We should point out that theoretical works concerning NTFPs often assume isotropic and homogeneous conditions because these premises make the problems more tractable. Still, we are not aware of any restrictions that prevent the occurrence of NTFPs in anisotropic non-homogeneous systems.

We observed universal behavior in an atomic superfluid driven far from equilibrium towards a turbulent state. Our work helps us to better understand a nonequilibrium-state evolution while in the vicinity of NTFPs, closely resembling recent results obtained in other out-of-equilibrium systems~\cite{Prufer2018,Erne2018,Eigen2018,Glidden2021,Galka2022}. However, our presented analysis shows the validity of this far-from-equilibrium theory beyond the limits explored up to this point in these other experiments. Our results are steps toward merging the quantum turbulence regime in trapped atomic gases into a class of systems that present dynamical universality by scaling. The obtained exponents may motivate future theoretical studies, and our cold-atom platform might be used to simulate different physical systems belonging to the same universality class.

\begin{acknowledgments}
We thank A.~Pi\~neiro Orioli, R.~P.~Smith, M.~Caracanhas, and T.~Gasenzer for fruitful discussions and G.D.~Telles for support with the experimental setup. This work was supported by the S\~ao Paulo Research Foundation (FAPESP) under the grants 2013/07276-1, 
2014/50857-8, 
2017/09390-7, 
and 2018/09191-7, 
and by the
National Council for Scientific and Technological Development (CNPq)
under the grants 465360/2014-9 and 
142436/2018-6. 
\end{acknowledgments}

\bibliography{ref.bib}

%

\end{document}